\documentclass[column,showpacs,preprintnumbers,amsmath,amssymb]{revtex4}
\preprint{}
\usepackage{graphicx}% Include figure files
\usepackage{dcolumn}% Align table columns on decimal point
\usepackage{bm}% bold math
\usepackage{mathrsfs}
\begin{document}
\title{Quantum estimation of detection efficiency with  no-knowledge quantum feedback}
\author{Dong  Xie}
\email{xiedong@mail.ustc.edu.cn}
\affiliation{Faculty of Science, Guilin University of Aerospace Technology, Guilin, Guangxi, P.R. China.}

\author{Chunling Xu}
\affiliation{Faculty of Science, Guilin University of Aerospace Technology, Guilin, Guangxi, P.R. China.}
\author{Jianyong Chen}
\affiliation{Faculty of Science, Guilin University of Aerospace Technology, Guilin, Guangxi, P.R. China.}
\author{An Min Wang}
\affiliation{Department of Modern Physics , University of Science and Technology of China, Hefei, Anhui, China.}

\begin{abstract}
We investigate that no-knowledge measurement-based  feedback control is utilized to obtain the  estimation  precision of  the
detection efficiency. For the feedback operators that concern us, no-knowledge measurement is the optimal way to estimate  the
detection efficiency. We show that the higher precision can be achieved for the lower or larger detection efficiency. It is found that no-knowledge feedback can be used to cancel decoherence. No-knowledge feedback with a high detection efficiency can perform well in estimating frequency and detection efficiency parameters simultaneously. And simultaneous estimation is better than independent estimation given by the same probes.
\end{abstract}

\maketitle

\section{Introduction}
Quantum metrology is a fundamental and important subject, which
concerns the estimation of parameters under the constraints of
quantum mechanics \cite{lab1,lab2,lab3}. There are widespread applications such as in timing, healthcare, defence, navigation,
and astronomy\cite{lab4,lab5,lab6,lab7,lab8}. For the mean-square error criterion, the Cram$\acute{e}$r-Rao
bound\cite{lab9,lab10,lab11} is the most well known analytic bound
on the error. The precision of the parameter is inversely proportional with quantum Fisher information(QFI).

The detection efficiency is a crucial physical quantity to judge the
detector quality. And it is
becoming more and more important to improve the estimation accuracy of  the detection efficiency\cite{lab12}.
Quantum feedback control\cite{lab13,lab14,lab15,lab16} can be used to improve the detection efficiency. In ref.\cite{lab12},  the use of the  different measurement-based quantum feedback types  to enhance the QFI
about the detection efficiency of the detector has been investigated.
In general, $\textbf{\textit{more signal, less noise}}$ from the detection can be used to better perform feedback control\cite{lab17,lab18}. However, Stuart S. Szigeti et al.\cite{lab188} show a significant surprising results: performing a no-knowledge measurement (no signal, only noise) can be advantageous in canceling decoherence. It is due to that a system undergoing no-knowledge monitoring has reversible noise, which can be canceled by directly feeding back the measurement signal. For a perfect detection efficiency $\eta=1$,  no-knowledge feedback control can be used to completely cancel  Markovian decoherence .

In this paper, we consider that the information of detection efficiency is encoded by a no-knowledge  measurement-based quantum feedback. When one uses the optimal feedback operator to cancel decoherence, no-knowledge quantum feedback control is the optimal way to measure the detection efficiency. For the low or high detection efficiency, the precision can be dramatically high. Finally, we show that simultaneous estimation of the frequency and detection efficiency parameters has an advantage over estimating the noise and phase parameters
individually, bringing us into the field of multi-parameter
quantum metrology\cite{lab19,lab20}. And no-knowledge quantum feedback can perform well in estimating the frequency with a high detection efficiency. Due to that the decoherence can be significantly canceled by the no-knowledge quantum feedback control with the high detection efficiency.

The rest of this article is arranged as follows. In Section II, we briefly introduce the quantum metrology of a single parameter and multi-parameter, and the formula of Fisher information. In Section III, we detail the physical model: a qubit interacting with a dephasing or dissipation reservoir and
 the feedback to the qubit by the no-knowledge homodyne measurement. Then, we obtain the precision of detection efficiency by the no-knowledge feedback in section IV. Simultaneous estimation of the frequency and detection efficiency parameters is studied in section V.  A conclusion and outlook are presented in Section VI.
\section{review of quantum metrology}
The famous Cram$\acute{e}$r-Rao bound\cite{lab9,lab10} offers a very good parameter estimation under the constraints of quantum physics:
\begin{eqnarray}
(\delta x)^2\geq\frac{1}{N\mathcal{F}_Q[\hat{\rho}_S(x)]},
\end{eqnarray}
where $N$ represents total number of experiments.  $\mathcal{F}_Q[\hat{\rho}_S(x)]$ denotes QFI, which can be generalized from classical Fisher information. The classical Fisher information is defined by
\begin{equation}
f(x)=\sum_k p_k(x)[d\ln[p_k(x)]/dx]^2,
\end{equation}
where $p_k(x)$ is the probability of obtaining the set of experimental results $k$ for the parameter value $x$. Furthermore,
the QFI is given by the maximum of the Fisher information over all measurement strategies allowed by quantum physics:
\begin{equation}
\mathcal {F}_{Q}[\hat{\rho}(x)]=\max_{\{\hat{E}_k\}}f[\hat{\rho}(x);\{\hat{E}_k\}],
\end{equation}
where positive operator-valued measure $\{\hat{E}_k\}$ represents a specific measurement device.

If the probe state is pure, $\hat{\rho}_S(x)=|\psi(x)\rangle\langle\psi(x)|$, the corresponding expression of QFI is
\begin{equation}
\mathcal {F}_{Q}[\hat{\rho}(x)]=4[\frac{d\langle\psi(x)|}{dx}\frac{d|\psi(x)\rangle}{dx}-|\frac{d\langle\psi(x)|}{dx}|\psi(x)\rangle|^2].
\end{equation}
If the probe state is mixed state, $\hat{\rho}(x)=\sum_k\lambda_k|k\rangle\langle k|$, the concrete form of QFI is given by
\begin{equation}
\mathcal {F}_{Q}[\hat{\rho}(x)]=\sum_{k,\lambda_k>0}\frac{(\partial_x \lambda_k)^2}{\lambda_k}+\sum_{k,k',\lambda_k+\lambda_k'>0}\frac{2 (\lambda_k- \lambda_k')^2}{\lambda_k+\lambda_k'}\langle k|\partial_x|k'\rangle.
\end{equation}
In general, it is complicated to calculate QFI. In this paper, we only consider two-dimensional system.  The QFI can be calculated explicitly by the following way\cite{lab21,lab22}
\begin{equation}
\mathcal {F}_{Q}[\hat{\rho}(x)]=\textmd{Tr}[(\partial_x\hat{\rho}(x))^2]+\frac{1}{\textmd{Det}(\hat{\rho}(x))}\textmd{Tr}[(\hat{\rho}(x)\partial_x\hat{\rho}(x))^2].
\end{equation}

For the classical multi-parameter Cram$\acute{e}$r-Rao bound:
\begin{equation}
 \textmd{Cov}(\widetilde{{\textbf{x}}})\geq F^{-1},
\end{equation}
where $\widetilde{{\textbf{x}}}=\{x_1,x_2, ..., x_m\}$, $\textmd{Cov}(\widetilde{\textbf{x}})$ refers to the covariance matrix for a locally
unbiased estimator $\widetilde{{\textbf{x}}}(k)$, $\textmd{Cov}(\widetilde{{\textbf{x}}})_{jk}=\langle(\widetilde{{x}}_j-x_j)(\widetilde{{x}}_k-x_k)\rangle$ and $\langle.\rangle$ represents the average with respect to the probability distribution $p_k(\textbf{x})$.
The classical Fisher matrix for $m$ parameters as the $m\times m$ matrix with entries given by
\begin{eqnarray}
F_{jk}=\sum_k p_k(x)\left(\frac{\partial \ln[p_k(x)]}{\partial x_j}\right)\left(\frac{\partial\ln[p_k(x)]}{\partial x_k}\right).
\end{eqnarray}
The multi-parameter QFI Cram$\acute{e}$r-Rao bound is given by
\begin{equation}
 \textmd{Cov}(\widetilde{{\textbf{x}}})\geq F_Q^{-1}, F_{Q_{ij}}=\frac{1}{2}\textmd{Tr}[\hat{\rho}(x)\{L_i,L_j\}],
\end{equation}
where the symmetric logarithmic derivative is obtained by
\begin{equation}
L_i=2\sum_{m,n}\frac{\langle \psi_m|\partial_{x_i}\hat{\rho}(x)| \psi_n\rangle}{p_m+p_n}|\psi_m\rangle\langle\psi_m|.
\end{equation}
Here, $p_{m,n}$ and $|\psi_{m,n}\rangle$ denote the eigenvalues and eigenvectors of density operator $\hat{\rho}(x)$.
For two-dimensional system, the multi-parameter QFI matrix is expressed by\cite{lab23}
\begin{equation}
 F_{Q_{ij}}=(\partial_{x_i}{\textbf{r}})\cdot(\partial_{x_j}{\textbf{r}})+\frac{(\textbf{r}\cdot\partial_{x_i}\textbf{r})(\textbf{r}\cdot{\partial_{x_j}\textbf{r})}}{1-|\textbf{r}|^2},
\end{equation}
where $\textbf{r}$ denotes the Bloch vector of $\hat{\rho}(x)$.

\section{a physical model of feedback control }
Consider a two-dimensional system with Hamiltonian $H$ interacts with a Markovian reservoir via the coupling operator $L$.
The system density operator, $\rho(t)$, evolves according to the master equation
\begin{equation}
\frac{d\rho(t)}{dt}=-i[H,\rho(t)]+\mathcal{D}[L]\rho(t),
\end{equation}
where $\mathcal{D}[L]\rho(t)=L\rho(t)L^\dagger-(L^\dagger L\rho(t)+\rho(t)L^\dagger L)/2$, and we have set $\hbar=1$ throughout the article.

For a homodyne measurement of the environment at angle
$\theta$, the conditional system dynamics are described by stochastic master equation(SME)\cite{lab24,lab25}
\begin{equation}
d\rho_c(t)=-i[H,\rho_c(t)]dt+\mathcal{D}[L]\rho_c(t)dt+\sqrt{\eta}dW(t)\mathcal{H}[Le^{i\theta}]\rho_c(t),
\end{equation}
where $dW(t)$  is the standard Wiener increment with mean zero and variance  $dt$. In the following, we consider  the  continuous  feedback control, and take the
Markovian feedback of the white-noise measurement record via a Hamiltonian. The continuous
measurement record can be described by the homodyne detection photocurrent\cite{lab26}
\begin{equation}
I(t)=\sqrt{\eta}\langle L e^{i\theta}+L^\dagger e^{-i\theta}\rangle_t+\xi(t),
\end{equation}
where $\xi(t)=\frac{dW(t)}{dt}$ is  Stratonovich noise.
When the coupling operator $L$ is Hermitian, such as dephasing in qubits ($L=\sigma_z$), no-knowledge measurement occurs for quadrature angle $\theta=\pm\pi/2$. The
measurement signal $I_{\pi/2}(t)=\xi(t)$ returns only noise.
 For non-Hermitian operator, such as dissipation reservoir ($L=\sigma_-$),  no-knowledge measurement requires an extra reservoir with coupling operator $L^\dagger$\cite{lab188}, giving the unconditional dynamics
 \begin{equation}
\frac{d\rho(t)}{dt}=-i[H,\rho(t)]+\mathcal{D}[L]\rho(t)+\mathcal{D}[L^\dagger]\rho(t).
\end{equation}
$\mathcal{D}[L]\rho(t)+\mathcal{D}[L^\dagger]\rho(t)=\mathcal{D}[L_+]\rho(t)+\mathcal{D}[L_-]\rho(t)$ with $L_\pm=i^{(1\mp1)/2}(L\pm L^\dagger)/\sqrt{2}$. Thus $L\pm$ are effective Hermitian coupling operators that admit no-knowledge measurements. The corresponding conditional system dynamics are described by stochastic master equation(SME)
\begin{equation}
d\rho_c(t)=-i[H,\rho_c(t)]dt+\mathcal{D}[L_+]\rho_c(t)dt+\mathcal{D}[L_-]\rho_c(t)dt+dW_+(t)\mathcal{H}[L_+e^{i\theta}]\rho_c(t)+dW_-(t)\mathcal{H}[L_-e^{i\theta}]\rho_c(t),
\end{equation}
where $dW_\pm(t)$ are independent Wiener increments.
Performing homodyne detection at two outputs, this yields the two corresponding measurement
signals
\begin{equation}
I_\pm(t)=2\cos \theta\sqrt{\eta}\langle L_\pm\rangle_t+\xi_\pm(t).
\end{equation}

\begin{figure}[h]
\includegraphics[scale=0.45]{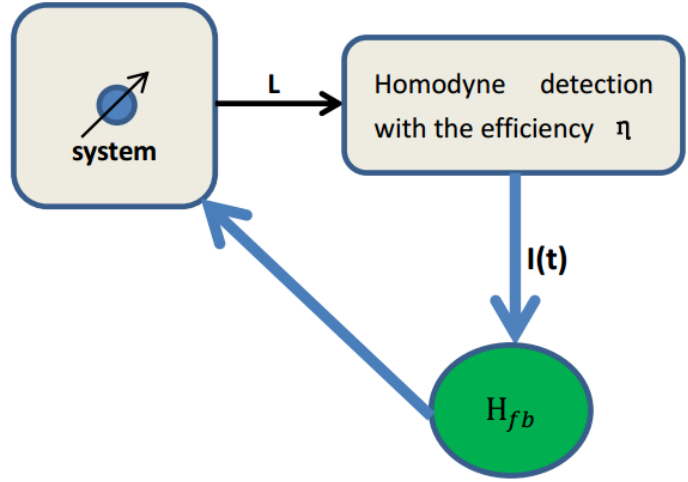}
 \caption{\label{fig.1} Schematic for a measurement-based feedback control protocol. Information about the system is
extracted by monitoring the decoherence channel $L$.  The
feedback Hamiltonian is applied to the system according to the  homodyne current $I(t)$ from the homodyne detection. Here, $\eta$ denotes the detection efficiency.}
 \end{figure}

Then the control Hermitian  can be written as $H_{fb}=I(t)F$ with $F$ is the feedback Hermitian operator, as shown in Fig. 1. For Hermitian coupling operator $L$, the stochastic equation for the conditioned
system state including feedback is:
\begin{equation}
d\rho_c(t)=-i[H,\rho_c(t)]dt+\mathcal{D}[L]\rho_c(t)dt-i\sqrt{\eta}[F,Le^{i\theta}\rho_c(t)+\rho_c(t)Le^{-i\theta}]dt+dW(t)\mathcal{H}[\sqrt{\eta}Le^{i\theta}-iF]\rho_c(t)+\mathcal{D}[F]\rho_c(t)dt.
\end{equation}
The unconditional master equation becomes\cite{lab25}
\begin{equation}
\frac{d\rho(t)}{dt}=-i[H,\rho(t)]+\mathcal{D}[L]\rho(t)-i\sqrt{\eta}[F,Le^{i\theta}\rho(t)+\rho(t)Le^{-i\theta}]+\mathcal{D}[F]\rho(t).
\end{equation}

For non-Hermitian operator $L$, the corresponding unconditional master equation is given by
\begin{eqnarray}
&&\frac{d\rho(t)}{dt}=-i[H,\rho(t)]+\mathcal{D}[L_+]\rho(t)-i\sqrt{\eta}[F_+,L_+e^{i\theta}\rho(t)+\rho(t)L_+e^{-i\theta}]\nonumber\\
&&+\mathcal{D}[F_+]\rho(t)+\mathcal{D}[L_-]\rho(t)-i\sqrt{\eta}[F_-,L_-e^{i\theta}\rho(t)+\rho(t)L_-e^{-i\theta}]+\mathcal{D}[F_-]\rho(t),
\end{eqnarray}
where $F_\pm$ denote independent feedback operators.
\section{the precision of detection efficiency }
For canceling decoherence, the optimal feedback operator is $F=L$ in the case of Hermitian operator $L$\cite{lab188}. In the case of non-Hermitian operator $L$, the optimal feedback operators are $F_\pm=L_\pm$\cite{lab188}. We consider that the feedback operator is proportional to the coupling operator: $F$=$\lambda L$ ($F_\pm$=$\lambda L_\pm$), where $\lambda$ denotes a real number.

Firstly, we consider the Markoivian dephasing operator $L=\frac{1}{\sqrt{2}}\sigma_z$. The Hamiltonian of system is given by $H=\frac{\omega}{2}\sigma_z$.
The master equation includes the coupling operator: $F$=$\lambda L$, which is given by
\begin{equation}
\frac{d\rho(t)}{dt}=-i[\frac{\omega}{2}\sigma_z,\rho(t)]+(1+\lambda^2-2\sin\theta\lambda\sqrt{\eta})\mathcal{D}[\frac{\sigma_z}{\sqrt{2}}]\rho(t).
\end{equation}
Choosing $|\varphi_0\rangle=\frac{1}{\sqrt{2}}(|0\rangle+|1\rangle)$ as the initial state, we can obtain the evolved density matrix
\[
\rho_\eta(t)= \frac{1 }{2}\left(
\begin{array}{ll}

 1, &e^{-i\omega t-(1+\lambda^2-2\sin\theta\lambda\sqrt{\eta})t}\\
  e^{i\omega t-(1+\lambda^2-2\sin\theta\lambda\sqrt{\eta})t}, &1\\
  \end{array}
  \right).
  \]\begin{equation}\end{equation}
    The QFI of $\eta$ is derived by Eq.(6),
    \begin{equation}
\mathcal {F}_{Q}[\rho_\eta(t)]=\frac{\sin^2\theta\lambda^2t^2e^{-2(1+\lambda^2-2\sin\theta\lambda\sqrt{\eta})t}}{\eta(1-e^{-2(1+\lambda^2-2\sin\theta\lambda\sqrt{\eta})t})}.
\end{equation}
From the above equation, we can see that for $\lambda>0$, $\sin\theta=1$ (or for $\lambda<0$, $\sin\theta=-1 )$ is chosen to obtain the maximal QFI.
 Namely, no-knowledge measurement ($\cos\theta=0$) is the optimal way. The optimal precision of detection efficiency $\eta$ is obtained by Eq.(1)
 \begin{eqnarray}
(\delta\eta)^2\geq\frac{\eta[1-e^{-2(1+\lambda^2-2\lambda\sqrt{\eta})t}]}{N\lambda^2t^2e^{-2(1+\lambda^2-2\lambda\sqrt{\eta})t}}.
\end{eqnarray}

Then, we can see that the uncertainty of $\eta$ can be 0 for $\eta=0$. And when $\lambda=1$, the uncertainty of $\eta$ can also be 0 for $\eta=1$.
Then, we can achieve the maximal QFI by taking the derivative of $t$ and $\lambda$. However, it is difficult to calculate the exact analytical solution. We make an approximation for $\eta<1$:
 \begin{eqnarray}
 \frac{\eta[1-e^{-2(1+\lambda^2-2\lambda\sqrt{\eta})t}]}{N\lambda^2t^2e^{-2(1+\lambda^2-2\lambda\sqrt{\eta})t}}\approx\frac{\eta}{N\lambda^2t^2e^{-2(1+\lambda^2-2\lambda\sqrt{\eta})t}}.
\end{eqnarray}
Then we can obtain the approximate optimal feedback constant $\lambda=1$ and the interrogation time
\begin{equation}
t_{opt}=\frac{1}{2-2\sqrt{\eta}}.
\end{equation}
Substituting them into Eq.(24), we can obtain the uncertainty of detection efficiency
 \begin{eqnarray}
(\delta\eta)^2\geq 4\eta(1-\sqrt{\eta})^2(e^2-1)/N.
\end{eqnarray}
A exact numerical solution is shown in Fig. 2. We can find that the approximate solution in Eq.(27) is close to the exact solution. Namely, Eq.(27) represents a good approximate analytical form.
\begin{figure}[h]
\includegraphics[scale=1]{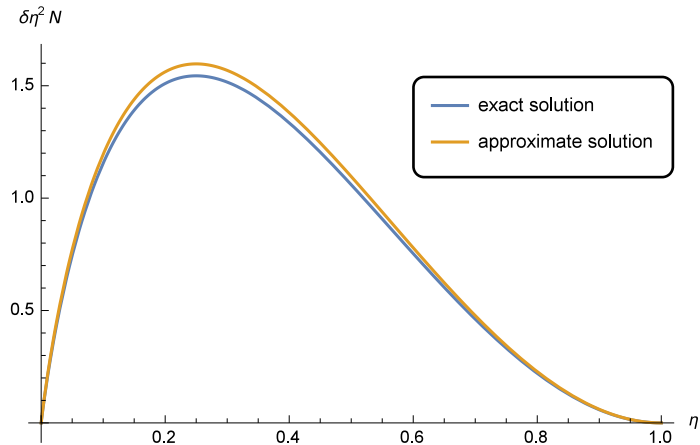}
 \caption{\label{fig.1} Diagram of the exact minimum precision and approximate analytical precision $N\delta\eta^2$ changing with the detection efficiency $\eta$.  The exact minimum precision comes from numerical calculation. The approximate analytical precision comes from Eq.(27).
It shows that the approximate analytical precision is close to the exact precision.}
 \end{figure}
And in general, the optimal feedback operator can be chosen to be the coupling operator ($\lambda=1$).

Then, we consider the Markoivian dissipative operator $L=\frac{1}{\sqrt{2}}\sigma_-$\cite{lab27}. \begin{equation}
\frac{d\rho(t)}{dt}=-i[\frac{\omega}{2}\sigma_z,\rho(t)]+(1+\lambda^2-2\sin\theta\lambda\sqrt{\eta})\mathcal{D}[\frac{\sigma_x}{2}]\rho(t)+(1+\lambda^2-2\sin\theta\lambda\sqrt{\eta})\mathcal{D}[\frac{\sigma_y}{2}]\rho(t).
\end{equation}
 Without loss of generality, it is simply to choose $\omega=0$. Choosing $\frac{1}{\sqrt{2}}(|0\rangle+|1\rangle)$ as the initial state, we can obtain the density matrix
 \[
\rho_\eta(t)= \frac{1 }{2}\left(
\begin{array}{ll}

 1, &e^{-i\omega t-(1+\lambda^2-2\sin\theta\lambda\sqrt{\eta})t}\\
  e^{i\omega t-(1+\lambda^2-2\sin\theta\lambda\sqrt{\eta})t}, &1\\
  \end{array}
  \right).
  \]
  So it is similar with the above case in estimating the detection efficiency.
  \section{Simultaneous estimation of the frequency and detection efficiency parameters}
 In general, simultaneous estimation of multi-parameter can perform better than estimating each parameter independently\cite{lab28}. We consider to  simultaneously estimate frequency $\omega$ and detection efficiency $\eta$. The information of $\omega$ and $\eta$ is encoded onto the evolved density matrix as shown in Eq.(22).

 Utilizing Eq.(7) and Eq.(11), we can achieve the precision of $\omega$ and $\eta$, which is given by

\[
\textmd{Cov}(\widetilde{\omega},\widetilde{\eta})\geq  \frac{1 }{t^2}\left(
\begin{array}{ll}

 e^{2(1+\lambda^2-2\sin\theta\lambda\sqrt{\eta})t}, &0\\
  0, &\frac{\eta [e^{2(1+\lambda^2-2\sin\theta\lambda\sqrt{\eta})t}-1]}{\lambda^2}\\
  \end{array}
  \right).
  \]\begin{eqnarray}\end{eqnarray}
  Considering the balanced importance of $\omega$ and $\eta$, the simultaneous estimation precision is given by
  \begin{eqnarray}
  \delta\omega^2+\delta\eta^2\geq \frac{1 }{t^2}\left[e^{2(1+\lambda^2-2\sin\theta\lambda\sqrt{\eta})t}+\frac{\eta [e^{2(1+\lambda^2-2\sin\theta\lambda\sqrt{\eta})t}-1]}{\lambda^2}\right].
  \end{eqnarray}
Obviously, no-knowledge measurement ($\sin\theta=1$) is the optimal way.
\begin{figure}[h]
\includegraphics[scale=1]{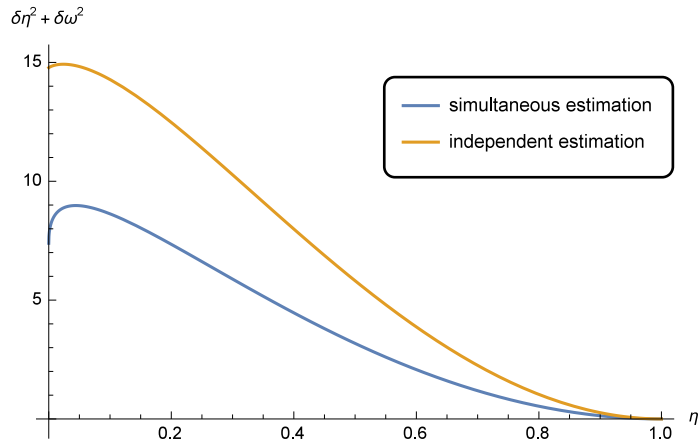}
 \caption{\label{fig.1} Diagram of simultaneous and independent estimation $\delta\eta^2+\delta\omega^2$ changing with detection efficiency $\eta$, given by the same probes. Here, we calculate the precision with an average of one probe.
It shows that the simultaneous estimation performs better than the independent estimation.}
 \end{figure}
Then, we numerically calculate the optimal precision of simultaneous and independent estimation with the same probes by choosing the optimal $t$ and $\lambda$, as shown in Fig. 3. The optimal $t$ and $\lambda$ can be different in  independently estimating $\omega$ and $\eta$, which is more flexible than the simultaneous estimation.  However, the independent estimation consumes twice the number of probes that the simultaneous estimation uses.
Hence, we can see that the simultaneous estimation can provide a better precision than the independent estimation. And the precision of estimating $\omega$ and $\eta$ is becoming smaller with the increase of detection efficiency.  It is because the no-knowledge feedback can better resist the decoherence with the larger detection efficiency.
\section{conclusion and outlook}
We have utilized the no-knowledge feedback control to estimate detection efficiency. The results show that when the feedback operator $F$ is proportional to  the coupling operator $L$, the no-knowledge measurement is the optimal way to encode the information of detection efficiency onto the probe state. By the exact numerical and approximate analytical calculation, we find  that the high precision of detection efficiency can be obtained for low or large detection efficiency. Finally, we show that a simultaneous estimation frequency and detection efficiency with no-knowledge feedback control can perform better than independent estimation.

Whether no-knowledge measurement is the optimal way for any feedback operator will be the further investigation. In this article, we only consider the Markovian feedback. Non-Markovian phenomenons are very common\cite{lab29,lab30}. Hence,  it is worth to utilize  non-Markovian feedback to estimate the precision of detection efficiency.

\section*{Acknowledgement}
 This research was supported by Guangxi Natural Science Foundation 2016GXNSFBA380227 and Guangxi Base Promotion Project of Young and Middle-aged Teachers(NO.2017KY0857).


\begin{thebibliography}{9}

\vspace{3mm}
\bibitem{lab1}Vittorio Giovannetti, Seth Lloyd, and Lorenzo Maccone, Phys. Rev. Lett. 96, 010401 (2006).
\bibitem{lab2}V. Giovanetti, S. Lloyd, L. Maccone, Science 306, 1330  (2004).
\bibitem{lab3}D. Xie and A. Wang, Phys. Lett. A 378, 2079 (2014).
\bibitem{lab4}]N.  Hinkley, J. A.Sherman, N. B. Phillips, M. Schioppo, N. D. Lemke, K. Beloy, M.
Pizzocaro, C. W. Oates, A. D. Ludlow, Science 341 (2013) 1215.
\bibitem{lab5}V. Giovannetti, S. Lloyd, and L. Maccone, Science 306, 1330
(2004); V. Giovannetti, S. Lloyd, and L. Maccone, Nat. Photon.
5, 222 (2011).
\bibitem{lab6}K. Bongs, R. Launay, and M. A. Kasevich, Appl. Phys. B 84,
599 (2006).
\bibitem{lab7}P. M. Carlton, J. Boulanger, C. Kervrann, J.-B. Sibarita, J.
Salamero, S. Gordon-Messer, D. Bressan, J. E. Haber, S. Haase,
L. Shao, et al., Proc. Natl. Acad. Sci. 107, 16016 (2010); M. A.
Taylor, J. Janousek, V. Daria, J. Knittel, B. Hage, H.-A. Bachor,
and W. P. Bowen, Nature Photon. 7, 229 (2013).
\bibitem{lab8}The LIGO Scientific Collaboration, Nat. Phys. 7 962 (2011); J.
Aasi et al. Nat. Photon. 7 613 (2013).
\bibitem{lab9} C. W. Helstrom, Quantum Detection and Estimation Theory, Academic, New
York, 1976.
\bibitem{lab10}S. L. Braunstein, C. M. Caves, Phys. Rev. Lett. 72, 3439 (1994).
\bibitem{lab11}V. Giovannetti, S. Lloyd, and L. Maccone, Science 306, 1330  (2004).
\bibitem{lab12}Shao-Qiang Ma, Han-Jie Zhu, and Guofeng Zhang, Phys. Lett. A, 381 1386 (2017).
 \bibitem{lab13}D. W. Berry, H. M. Wiseman, Phys. Rev. A 65, 043803 (2002).
 \bibitem{lab14}H. M. Wiseman, Phys. Rev. A 49, 5159 (1994).
 \bibitem{lab15}H. M. Wiseman, and G. J. Milburn, Phys. Rev. Lett. 70, 548 (1993) .
\bibitem{lab16}Naoki Yamamoto, Phys. Rev. A 72, 024104 (2005).
\bibitem{lab17}M. O. Scully and M. S. Zubairy, Quantum Optics, 1st ed.
(Cambridge University Press, 1997).
\bibitem{lab18}N. P. Robins, P. A. Altin, J. E. Debs, and J. D. Close,
Atom lasers: production, properties and prospects for precision inertial measurement, Physics Reports 529, 265
(2013).
\bibitem{lab188}S. S. Szigeti, A. R. R. Carvalho, J. G. Morley and M. R. Hush, Phys. Rev. Lett. 113, 020407 (2014) .
 \bibitem{lab19}M. Szczykulska, T. Baumgratz, and A. Datta, Adv. Phys.: X 1,
621 (2016).
\bibitem{lab20}S. Ragy, M. Jarzyna, and R. Demkowicz-Dobrzanski, Phys. ¡ä
Rev. A 94, 052108 (2016).
\bibitem{lab21}J. Dittmann, J. Phys. A 32 (1999) 2663.
\bibitem{lab22}W. Zhong, Z. Sun, J. Ma, X. G. Wang, F. Nori, Phys. Rev. A 87 (2013) 022337.
\bibitem{lab23}Z. Jiang, Phys. Rev. A 89, 032128 (2014).
\bibitem{lab24}H. M. Wiseman and G. J. Milburn, Quantum Measurement and Control (Cambridge University Press, 2010).
\bibitem{lab25} V. B. Braginsky and F. Y. Khalili, Rev. Mod. Phys. 68,
1 (1996).
\bibitem{lab26}H. M. Wiseman and G. J. Milburn, Phys. Rev. A 47 (1993) 642.
\bibitem{lab27} A. R. R. Carvalho and M. F. Santos, New Journal of
Physics 13, 013010 (2011).
\bibitem{lab28}Lu Zhang, Kam Wai Clifford Chan, Phys. Rev. A 95, 032321 (2017).
\bibitem{lab29}Marco Cianciaruso, Sabrina Maniscalco, Gerardo Adesso, Phys. Rev. A 96, 012105 (2017).
\bibitem{lab30}Aniello Lampo, Jan Tuziemski, Maciej Lewenstein, Jaroslaw K. Korbicz, Phys. Rev. A 96, 012120 (2017).
\end{thebibliography}
\end{document}